\RequirePackage{fix-cm}
\documentclass{svjour3}                     
\smartqed  
\usepackage{graphicx}

\usepackage{breqn} 
\usepackage{supertabular} 
\usepackage{natbib}
\usepackage[colorlinks=true,allcolors=blue]{hyperref}

\journalname{Experiments in Fluids}
\begin{document}

\title{Control of long-range correlations in turbulence}


\author{Kevin P. Griffin$^{*}$\thanks{\noindent$^{*}$These authors contributed equally: N. J. Wei and K. Griffin} \and Nathaniel J. Wei$^{*}$ \and Eberhard Bodenschatz \and Gregory P. Bewley$^{\dagger}$\thanks{\noindent$^{\dagger}$Corresponding author}}


\authorrunning{K. P. Griffin, N. J. Wei, E. Bodenschatz, \& G. P. Bewley} 

\institute{ Kevin Griffin \at
              Stanford University; Stanford, CA 94305 \\
              \email{kevinpg@stanford.edu}
            \and
            Nathaniel Wei \at
              Princeton University; Princeton, NJ 08544 \\
              \email{njwei@stanford.edu}
            \and
            Eberhard Bodenschatz \at
              Max Planck Institute for Dynamics and Self-Organization; G{\"o}ttingen, Germany 37077 \\
              \email{eberhard.bodenschatz@ds.mpg.de}
            \and
            Gregory Bewley \at
              Cornell University; Ithaca, NY 14850 \\
              \email{gpb1@cornell.edu}
}

\date{Received: date / Accepted: date}

\maketitle

\begin{abstract}
\label{abstract}

\qquad The character of turbulence depends on where it develops.  
Turbulence near boundaries, for instance, is different than in a free stream.  
To elucidate the differences between flows, it is instructive to vary the structure of turbulence systematically, 
but there are few ways of stirring turbulence that make this possible.  
In other words, an experiment typically examines either a boundary layer or a free stream, say, 
and the structure of the turbulence is fixed by the geometry of the experiment.  
We introduce a new active grid with many more degrees of freedom than previous active grids.  
The additional degrees of freedom make it possible to control various properties of the turbulence.  
We show how long-range correlations in the turbulent velocity fluctuations can be shaped 
by changing the way the active grid moves.
Specifically, we show how not only the correlation length but also the detailed shape of the correlation function 
depends on the correlations imposed in the motions of the grid.  
Until now, large-scale structure had not been adjustable in experiments.  
This new capability makes possible new systematic investigations into turbulence dissipation and dispersion, 
for example, and perhaps in flows that mimic features of boundary layers, free streams, 
and flows of intermediate character.  

\keywords{active grid \and energy decay \and anisotropy \and correlation function \and integral length scale \and isotropy}
\PACS{47.27.Rc turbulence control \and 47.27.wb turbulent wakes}
\subclass{76F70 control of turbulent flows}
\end{abstract}


\section{Introduction}
\label{intro}

Active grids, or grids with moving parts called winglets, 
were invented to produce higher turbulence intensities in wind tunnels than had been possible with classical grids, 
which were composed of fixed bars. Classical grids provided a platform for studying nearly isotropic turbulence \cite[e.g.][]{Corrsin1942, comte-bellot}. Pioneering active grid work used jets to increase turbulence intensity \cite[][]{Mathieu1965,Gad-el-Hak1974}, with recent work improving the homogeneity and isotropy of the flow \cite[][]{Szaszak}. \cite{makita} demonstrated that active grids composed of rotating winglets could both achieve higher Reynolds numbers and afford better flow homogeneity than classical grids.  

Wind tunnel turbulence can be made homogeneous more easily than other types of flows, which can help reveal universal aspects of turbulence.  
Increasing the Reynolds number of the turbulence can be similarly enlightening.  
Though the large scales of turbulence are not universal 
\cite[e.g.][]{Blum2011}, 
the smalls scales may be \cite[][]{Kolmogorov1941b}.  
The art of wind-tunnel experiments has therefore been to increase the turbulence intensity, and so the Reynolds number, without spoiling the homogeneity of the turbulence, thus revealing the universal character of the small scales.  
An active grid can simultaneously fulfill these objectives; the pioneering work of \cite{makita} has been furthered by many subsequent investigations \cite[e.g.][]{mydlarski,kang2,cekli2,knebel2,thormann}.  

Active grids promise not only to produce higher Reynolds numbers than other grids 
but also to afford additional measures of control over the properties of the turbulence.  
Because active grids have moving parts, and thus a variable geometry, 
the positions of the different parts and the way they move relative to each other 
determine the patterns in the flow through the grid.  
Deliberately varying the active-grid geometry and motion 
may make it possible to vary not only the intensity of the turbulence 
but also the extent to which fluid motions are correlated across the tunnel, 
the degree to which the turbulence is homogeneous or inhomogeneous, 
and the degree to which it is isotropic or anisotropic.  
Many of these opportunities have been explored and reported in the literature, 
and a review of some of this work appears in \cite{mydlarski_review}. 

These advances in active grid design have led to progress in various multi-physics and application areas such as particle transport in turbulence \cite[][]{Ayyalasomayajula2006,Saw2012,shen3,Siebert2010,Obligado2015,Gerashchenko2013}, turbulence with a mean temperature gradient \cite[][]{Gylfason2004}, and the atmospheric boundary layer interacting with a wind turbine array \cite[][]{Cal2010,wachter,Maldonado2015}. Precise recreation of the atmospheric boundary layer requires that the experimenter control the mean profile across the wind tunnel. \cite{Zhou2019} have computationally shown that a sequence of stationary cylinders can be used to tailor the profile of the mean flow. \cite{Hearst2017} have used an active grid to create various mean flow profiles. Many other contributions are reviewed in \cite{mydlarski_review}. 

In these previous studies, the active grids were composed of rows and columns of winglets which were attached to the same axis and moved together.  
Typically there are about 10 rows and 10 columns, 
for a total of about 20 degrees of freedom for the grid 
\cite[][]{shen1,kang,knebel,cekli3,varshney,wachter,weitemeyer,hearst,thormann2}. This limitation exists even for very large active grids, such as that of \cite{Larssen2011}, which spans 2.14m and has only 20 degrees of freedom.
In contrast to these designs, the grid we designed and built has an order of magnitude more degrees of freedom.  
This increase was achieved by using 129 individual servo motors -- one for each winglet -- 
instead of a single motor to control each row or column of winglets.  

In this paper, we first describe this new type of active grid.  
We then explain how we varied turbulence properties by programming the motions of the grid.  
In some cases our results are more convincing than in others; 
in all cases we believe they suggest interesting future work.  
For instance, we found that increasing the amplitude of the motions of the winglets 
causes the turbulence intensity of the flow to increase.  
While this result is more or less intuitive, 
it may be less obvious that adding correlations to the random motions of the winglets 
causes the turbulent fluid motions to be more correlated with each other, even far downstream of the grid.  
We also show how the turbulence can be made more or less isotropic 
by varying the balance between spatial and temporal correlations in the motions of the winglets.  
Substantial control of isotropy has so far been reported only in \cite{chang} and \cite{bewley} 
using a soccer-ball apparatus with a negligible mean flow.

\subsection{List of Symbols}
\noindent \begin{supertabular}{p{0.18\linewidth}p{0.75\linewidth}}
$u,v$& Streamwise and transverse velocities\\
$u^{\prime}$& Root mean square (RMS) of the velocity fluctuations $u-U$\\
$U$& Mean of the streamwise velocity $u$\\
$T^*$& Ratio between time scales of grid and fluid motions \\
$\theta(t,y,z)$& Angle of a winglet at time $t$ and position $(y, z)$ on the grid\\
$u_{\mathrm{tip}}$& Tangential velocity of the tip of a winglet \\
$R_{\mathrm{grid}}(t,y,z)$& Winglet angle autocorrelation\\
$l_{\mathrm{grid}}$&Characteristic length scale of the winglet autocorrelation\\
$M$& Grid spacing, 115 mm\\
$r$&Correlation distance in the $x$ (streamwise) direction\\
$R_{uu}(r)$&$u$-velocity autocorrelation\\
$L$&Integral length scale, computed from $R_{uu}$\\
$\lambda$&Taylor microscale\\
$\nu$&Kinematic viscosity of air\\
${\mathrm{Re}}_{\lambda}$&Turbulence Reynolds number, $ u^{\prime} \lambda / \nu$\\
$\zeta_{\mathrm{grid}},\zeta_{\mathrm{flow}}$& Shape parameters for characterizing correlation functions of the grid and of the flow\\
$I_{\mathrm{grid}}$&Winglet anisotropy coefficient, ratio of the sizes of the winglet correlation kernels in the spatial dimension over that of the temporal dimension\\
$x_0$& Virtual origin of energy-decay power-law fit\\
\end{supertabular}

\section{Experimental Setup}
\label{exp_setup}

The design of the active grid detailed in this work is addressed in this section. First, since the grid was intended for use in a pressurized sulfur-hexafluoride tunnel, a detailed description of the considerations behind the construction of the grid is given. Then, the control algorithms developed for generating turbulence with the grid are explained. In order to demonstrate the turbulence-control capabilities of the active grid and its control protocols, the experiments presented in this paper were conducted in another wind tunnel. Thus, the setup and instrumentation used in this wind tunnel are documented as well.

\subsection{Mechanical design of the active grid}
\label{sec:mech_design}

\begin{figure}
        \includegraphics[width=\columnwidth]{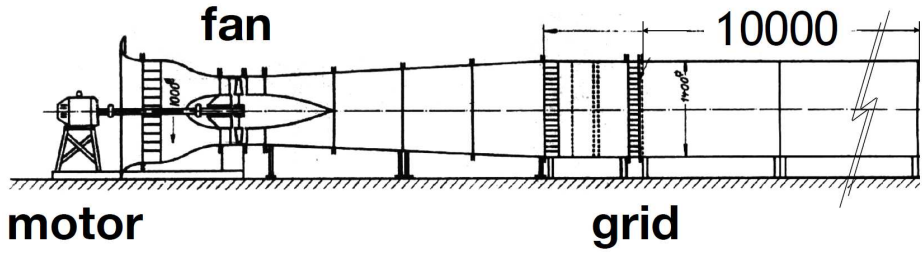}
        \caption{The open wind tunnel is composed of a fan, conditioning honeycombs, the active grid, and a 9.4-m test section.}
        \label{fig:tunnel_schematic}
\end{figure}

Though the results presented in this paper were obtained in a standard open-return wind tunnel, we designed our grid to operate in the Variable Density and Turbulence Tunnel (VDTT) \cite[][]{bodenschatz}, which imposed significant constraints on the design of the grid.  
The VDTT circulates pressurized sulfur hexafluoride, which is heavy 
-- its mass density at 15 bar of 0.1\,$\mathrm{g/cm^3}$ is closer to water than to air.  
Though the tunnel is a low-speed tunnel, operating at up to 5\,$\mathrm{m/s}$, 
the high density produces large forces on the grid and large torques on the servo motors 
relative to those experienced by an active grid in air.  
The tests presented in this paper, however, were not performed in the VDTT.  
The grid was instead installed in the Prandtl tunnel \cite[see][]{Reichardt1938, Eckelmann1983}, shown here in Fig.\ \ref{fig:tunnel_schematic}. 
The objectives of our active-grid design were to generate high Reynolds-number turbulence, 
to be sufficiently durable that all desired experiments could be performed, 
to produce turbulence that could be approximately homogeneous and isotropic if desired, 
and to allow variation of the turbulence properties 
-- to make the turbulence inhomogeneous and anisotropic, for instance.  

Active grids are composed of moving winglets. In principle, any tessellation of the plane can serve as a pattern for winglet shapes.  
These include squares or hexagons rotating about any of their symmetry axes, or other parallelepipeds.  
The active grid shown in \cite{makita} was composed of diamond-shaped winglets.  
\cite{thormann} observed the effects on the flow of different fractal shapes 
inscribed within the diamond-shaped outline.  
We employed the simple solid diamond shape as in the original Makita design 
because it is simple and we did not know of any evidence, and could not think of any argument, 
for why the outline of the grid winglets should substantially influence the properties of the resultant turbulence.  

There are two competing design objectives regarding turbulence length scales.  
First, in order to produce the highest Reynolds numbers, 
the scale at which the grid produces turbulence needs to be as large as possible.  
For the VDTT, the maximum length is given by the width of the wind tunnel, 1.5\,m.  
The second objective, in contrast, is to produce a uniform flow in the center of the tunnel 
that evolves without interacting with the boundaries, 
which requires that the scale of the turbulence be much smaller than the width of the tunnel.  
These competing design objectives motivated our design, 
which allows a user to specify the extent to which motion of the winglets are correlated.  
The idea is that a user can vary the length scale of these correlations 
to generate turbulence at a range of prescribed scales.  

According to the standard for classical grids, 
the spacing of the grid should be at least ten times smaller than the width of the wind tunnel.  
Under this constraint, the grid spacing was optimized to minimize the difference 
between the dimensions of the array of winglets and each dimension of the octagonal test section in the VDTT.  
There was no exact solution, 
but among several minima a spacing of $M$ = 115\,mm was chosen somewhat arbitrarily.  
Dividing the width and height of the tunnel by this dimension 
results in an array of winglets with 11 rows and 13 columns. 

\begin{figure}
        \includegraphics[width=\columnwidth]{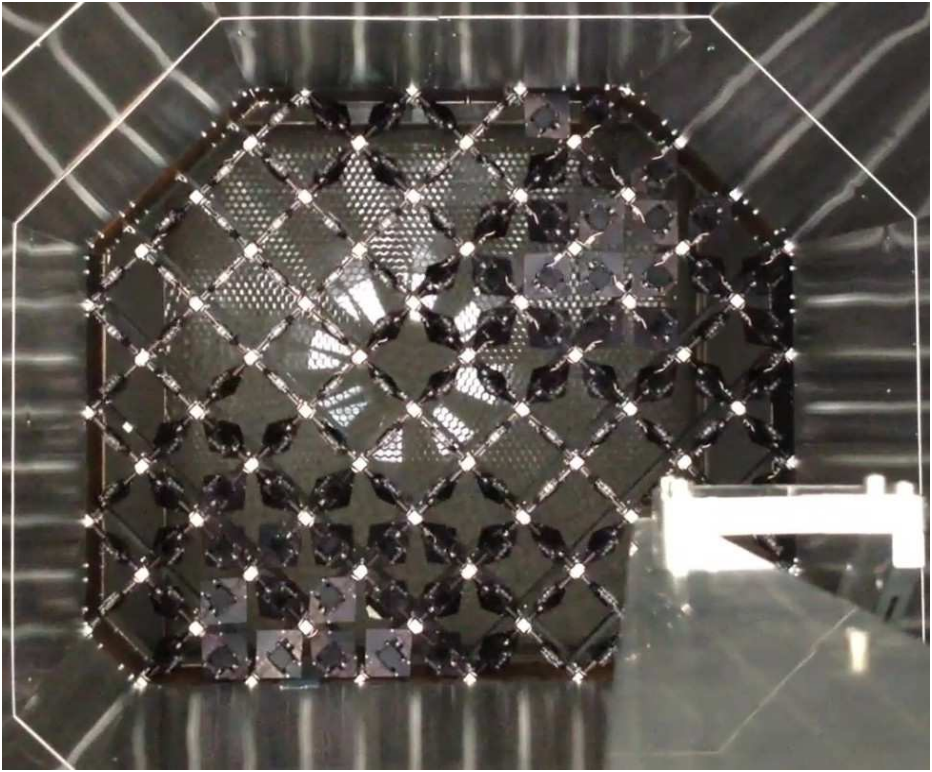}
        \caption{The active grid viewed from downstream, inside the octagonal test section. 129 of the winglets shown in Fig.\ \ref{fig:paddle} form an octagonal grid. The octagonal section profile was chosen to match that of the VDTT. In this image, the position of the winglets are spatially correlated: closed winglets tend to be near other closed winglets, and open winglets tend to be near other open ones. The velocity-probe holder is visible in the foreground of this image (lower right). The width and height of the test section are 1.5~m and 1.27~m, respectively.}
        \label{fig:grid_lighted}
\end{figure}

The test section of the VDTT has an unusual eight-sided cross section\footnote{An eight-sided cross section 
was chosen by the designer 
to maximize the area available to the flow within the circular cross section of the VDTT pressure vessel, 
while allowing equipment to be installed between the wall of the test section and the wall of the pressure vessel} 
that is 1.5\,m wide and 1.3\,m high, as illustrated in Fig.\ \ref{fig:grid_lighted}.  
The octagonal shape requires that the corners of the $13 \times 11$ array be omitted, 
so that there are 129 winglets as shown in Fig.\ \ref{fig:grid_lighted}.  

On most previous grids the rotation of the winglet axes were speed-controlled, 
with the direction of rotation changing at random intervals.  
Such grids are easier to design and build.  
Our grid, like the one described in \cite{cekli}, is position-controlled.  
\cite{cekli} shows how position control can shape the mean velocity profile 
to keep the velocity constant across the width of the tunnel or to produce a mean shear as desired.  
In our grid, we take advantage of position control 
to adjust the positions of winglets with respect to each other.  
Position control also enables our motion-control algorithms to adjust the blockage 
that the grid presents to the oncoming flow.  

An active grid can be thought of as an intricate butterfly valve.  
If the winglets move collectively from the open to the closed position while the mean wind speed is fixed, 
the grid generates a larger pressure drop.  
This larger pressure drop corresponds to both higher turbulence intensities 
and to a larger force needed to hold the grid in place.  
This force can be estimated by treating the grid analytically as a perforated plate when it is closed, 
as well as by reviewing the performance of commercial butterfly valves and perforated plates.  
For our grid, the maximum blockage of 80\% was selected so that the force generated by the grid, 
about 40\,kN, would not break the wind tunnel, 
as determined by computer modeling of the stresses in the wind tunnel structure.

To select appropriate servo motors, 
we needed to estimate the aerodynamic torque on each winglet at different angles with respect to the flow.  
The winglets were diamond-shaped with a diagonal of 150\,mm and were rotated about one of their diagonals on an axle.  
Thus, each winglet can be thought of as a delta-shaped wing 
that produces lift at small angles of attack and stalls as the angle is increased.  
Dimensional arguments produce an estimate for the maximum torque on the winglet: 
$\tau_{max} = C_\tau \rho U^2 M^3$, which is reached just before the winglet stalls.  
Here, $C_\tau$ is a dimensionless coefficient, $\rho$ is the density of the air, 
$U$ is the mean speed of the wind, and $M$ is the spacing of the grid.  
An estimate of the RMS torque experienced by the winglet as it rotates 
is about $2 / 3$ of the maximum torque, 
under the assumption that the winglet visits every angular position with equal probability.  

Simple experiments indicated that the torque coefficient, $C_\tau$, 
was about 0.15 and was constant over the range of winglet Reynolds numbers 
between about 20,000 and 120,000 that we tested.  
The tests were performed in the Warhaft Wind and Turbulence Tunnel at Cornell University \cite[][]{yoon}.  
To do the experiments, we measured the torque on an individual winglet.  
The torque was measured by resting a lever arm connected to the axle on a digital scale.  
We measured the torque on the winglet for various angles of the winglet with respect to the flow, 
and for various flow speeds.  
At each flow speed, the torque on the winglet increased steadily with its angle relative to the flow from zero 
until the winglet stalled.  
After stalling, the torque decreased to zero again until the winglet was perpendicular to the flow, as expected.  

An active grid produces turbulence when it is static, 
but the motions of the winglets also contribute to the turbulence.  
Indeed, \cite{thormann} used winglets of different shapes and sizes at different positions on the grid 
to show how a gradient across the tunnel in the number and size of the winglets 
can introduce a shear-free gradient in the turbulence intensity downstream of the grid.  

A time scale characterizing the grid motion is the time $T_g$
it takes for a winglet to move from a position parallel to the flow to one where it is blocking the flow.  
This is a measure of the time it takes for the grid to change shape, as seen by the flow.  
The characteristic time of the energy-containing scales of the turbulence is given by $T_L = L/U$, 
where $L$ is the correlation length associated with $u^{\prime}$.  
We compare the two time scales by their ratio $T^* = T_g / T_L$.  
Similar parameters have been proposed in \cite{shen1}, in \cite{mydlarski_review}, and in \cite{poorte}. 

In the case that $T^* > 1$, the grid moves more slowly than the turbulence, 
and the resulting turbulence evolves through a series of a quasi-steady states 
determined by the set of the winglet positions.  
When $T^* < 1$, the grid is faster than the turbulence, 
and the flow might not respond to the instantaneous spatial structures imprinted by the winglet configurations.  
In this regime, the correlation length of the flow is probably given by the winglet size, 
and not by the correlation length in the winglet positions.  
Therefore, we argue that if we wish to dynamically control the forcing scale of the turbulence, 
then the ratio of the grid time scale to the flow time scale should be close to one.  
To explore this relationship between $T^*$ and the turbulence in experiments, then, we would like to achieve values of $T^*$ significantly less than one -- that is, at the maximum rotation rate of the winglets,
the grid should be able to respond more quickly than the flow.
The active grids used in \cite{garg}, \cite{thormann}, and \cite{makita} 
are characterized by time scales comparable to the large eddy turnover time, $T_L$.  
The grid in the Netherlands is about ten times faster \cite[][]{cekli}.  
One purpose of the data we present below is to support these arguments.

On the basis of the above calculations and tests, 
we required servos that could work against an RMS torque of about 0.5\,Nm.  
We identified servos on the hobby market with several advantages: 
small packaging, so that the servos would fit within the outline of a winglet; 
a computer internal to each servo to enable position control; 
and prices far lower than for specialized industrial servo motors 
due to the economies of scale as mass-produced consumer products.  

\begin{figure}
\centering
        \includegraphics[width=\columnwidth]{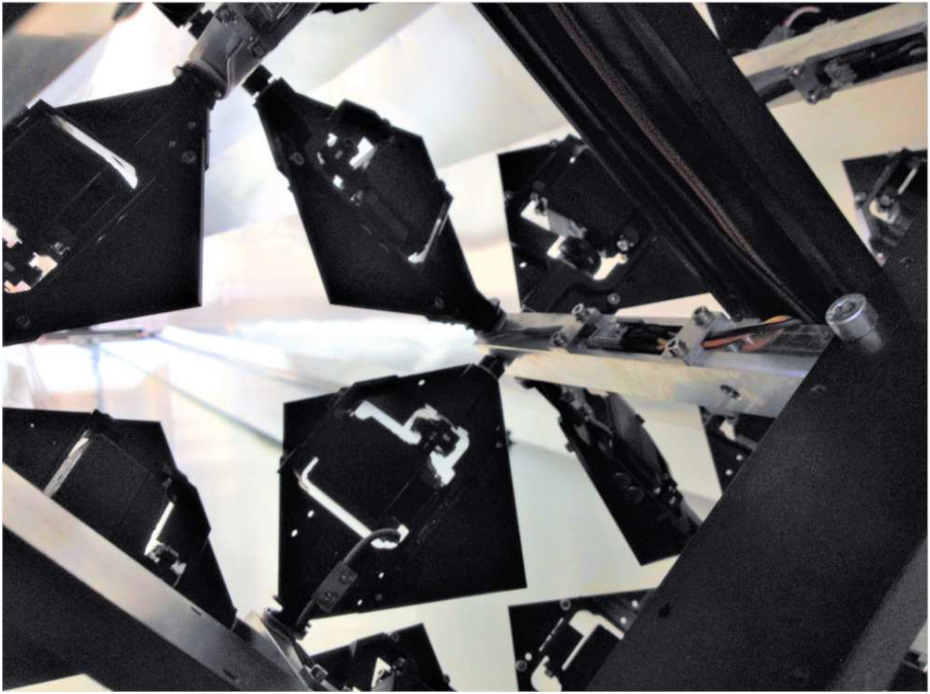}
        \caption{Several of the 129 winglets in the grid are shown.  Each winglet (square plate) has a servo embedded in it. This allows the position of each winglet to be controlled independently. The diagonal dimension of each winglet is 15 cm. }
        \label{fig:paddle}
\end{figure}

As Fig.\ \ref{fig:paddle} shows, each winglet was square and rotated on its diagonal 
up to 90$^\circ$ in either direction from the open orientation. A 3D rendering of the servo-winglet assembly is given in Fig.\ 21 of \cite{bodenschatz}.
The winglets could rotate 90$^\circ$ in about 0.23\,s.  

We tested several hobby servos from each major manufacturer.  
The servos were placed in a jig and made to work against a coil spring 
that supplied the equivalent of the anticipated maximum aerodynamic torque.  
The servos pulled against the spring about twice per second repeatedly, until the servos failed.  
The Futaba BLS152 servo motors survived far longer than any other servo on the market.  
The tests, however, revealed that when the BLS152 servos did fail, 
they failed where plastic gears on the high-speed side of the gear train turned low-speed gears made of metal.  
The plastic gears in the gear train were therefore replaced by hand with metal gears supplied by Futaba.  

The active grid was supported by a backbone composed of a lattice of steel channels, 
seen as diagonals in Fig.\ \ref{fig:grid_lighted} and Fig.\ \ref{fig:paddle}.  
The lattice forms a classical grid with negligible blockage compared to the active grid it supports, as the projected area of the winglets exceeds that of the lattice even at small winglet angles.  
The channels had covers and served as conduits for the wires that powered the servos 
and for those that carried the position signals.  
The steel lattice also supported an array of posts; each held the axles of four winglets.  
The array of winglets was about 15\,cm downstream of the steel lattice.  

The winglets consisted of a cage that supported the servo, 
so that each servo rotated along with its own winglet as shown in Fig.\ \ref{fig:paddle}.  
The servo cage was designed to be the minimal structure mechanically necessary, 
and was made of CNC-machined aluminum.  
Fins could be attached to make the winglet diamond shaped, 
as in the original Makita grid; 
fins with other shapes could also be attached but have not yet been tested.  
No effort was made to make the winglet aerodynamic or streamlined, 
as the purpose of the grid was to produce turbulence.  
The winglets were, however, made as thin as possible 
in order to maximize the difference of their effect on the flow between their open ($0^\circ$) 
and closed ($\pm 90^\circ$) positions.  
This winglet structure was tested for strength with static weight 
far in excess of the expected aerodynamic pressure and did not fail.

\subsection{Electrical design of the active grid}

The grid was powered by a rack of power supplies operating at 6\,V, 
with a total of about 6\,kW of available power.  
Each servo had its own dedicated 7.5\,A fuse between it and its power supply.  
To simplify the design and avoid the use of slip rings, 
the wires for each servo were wrapped around the axle of the winglet about four times.  
As the winglets rotated, the wires wrapped and unwrapped.  

The servos responded to pulse-width modulated 
signals 
synthesized by two SD84 servo control boards manufactured by Devantech Limited.  
The SD84 boards, in turn, took instructions via USB from an Apple MacBook, 
for which we developed computer software that we describe below.

\subsection{Principles of the control algorithm}
\label{sec:grid_correlations}

The motions of the individual winglets in our grid were correlated with each other, 
and the properties of the correlations could be varied by the user.  
Correlations between the winglets were introduced by convolving a kernel 
with an array of randomly-generated angular positions, one for each winglet on the grid.  
This procedure produced constantly changing but correlated winglet positions. In more detail, a three-dimensional matrix $A$ of random winglet angles 
was convolved with a correlation kernel $K$ in order to calculate a set of angles, $\theta$, 
so that 
\begin{dmath} \label{eqn:corr}
\theta(t,y,z) = \\ 
\int K(t^{\prime}, y^{\prime}, z^{\prime}) A(t-t^{\prime}, y-y^{\prime}, z-z^{\prime}) dt^{\prime} dy^{\prime} dz^{\prime}.  
\end{dmath}
Here, $y$ and $z$ correspond to orthogonal directions in the plane of the grid, 
and $t$ corresponds to time.  
In practice, the angles for each winglet were spaced in time by 0.1\,s.  
Fig.\ \ref{fig:grid_lighted} shows a picture of the grid in the midst of motion produced by this method, 
with regions of highly correlated winglets readily apparent throughout. More views are given in Fig.\ 21 of \cite{bodenschatz}. 

The correlations between winglets as a function of displacements 
along the streamwise and normal directions, with respect to a given reference winglet located at $(t_0,y_0,z_0)$, 
could be calculated according to  
\begin{equation}
R_{\mathrm{grid}}(t, y, z) = \frac{\langle \theta(t, y, z)\theta(t_0, y_0, z_0) \rangle}{\langle \theta(t_0, y_0, z_0)^2 \rangle}, 
\end{equation}
where $\theta$ is the angular position of a winglet and $R_{\mathrm{grid}}$ is the normalized correlation.  
The angle brackets indicate a time average over the duration of an experiment.

The size and shape of the kernel, $K$, 
determines the shape of the winglet correlation function, $R_{\mathrm{grid}}$.  
In Sec.~\ref{Results} we demonstrate that varying the shape and size of the winglet correlation functions, 
that is, varying the resulting grid motion, alters various turbulence properties.  

\begin{figure}
        \includegraphics[width=\columnwidth]{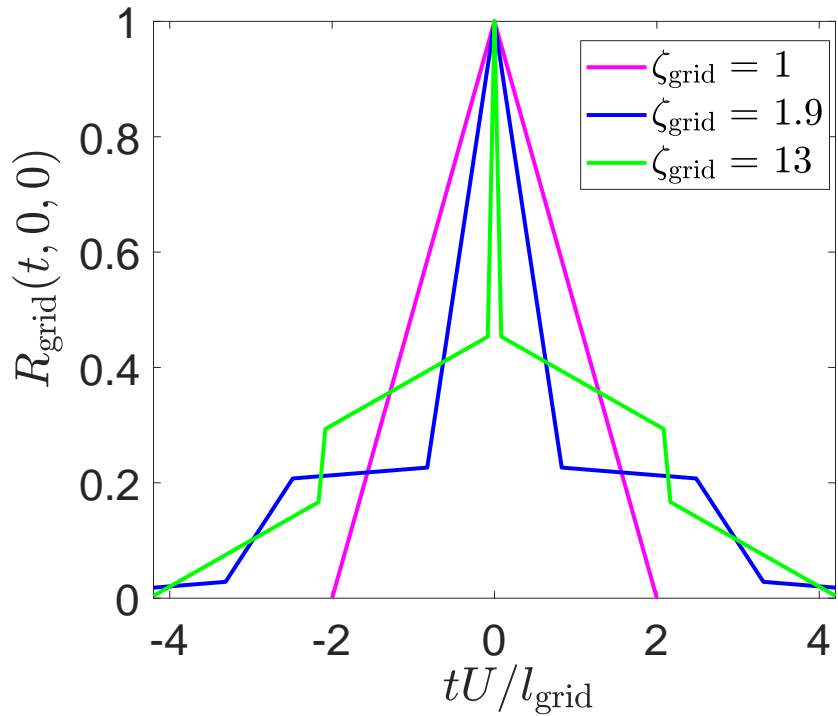}
        \caption{This plot shows the correlations between the angular displacement of an arbitrary winglet and its future and past displacements. The three curves illustrate that the shape of such winglet correlations can be controlled.}
        \label{fig:paddle_corr}
\end{figure}

We used two different classes of kernels to generate two types of winglet correlation functions.  
One kernel was a simple boxcar with a certain width, which we called a ``top hat,''
and the other was a superposition of two boxcars with different widths and heights, 
which we called a ``long tail'' because it consisted of a narrow boxcar filter with filter height 1 superimposed on a wide boxcar with filter height $<1$. This wide boxcar is referred to as the tail of the correlation.
Fig.\ \ref{fig:paddle_corr} shows the winglet correlation shapes that result from the autocorrelation of the boxcar (pink) and the long tail (green and blue) kernels. The presence of the tail in the long tail leads to a steeper slope near the reference winglet and a more gradual slope at large correlation distances. These kernels were applied to the streamwise ($t$) and normal ($y$ and $z$) components of the flow with separate correlation distances, so that the isotropy of the grid correlations (i.e.\ streamwise vs.\ normal correlation lengths) could be controlled.

The wide range of correlations that the active grid could produce 
necessitates the definition of parameters for describing the shape and size of these winglet correlations. 

The characteristic length scale of the winglet correlation is given by an integral length scale, which is defined as
\begin{equation}
l_{\mathrm{grid}} = U \int_0^{\infty} R_{\mathrm{grid}}(t, 0, 0)\, dt. 
\end{equation}
$l_{\mathrm{grid}}$ is referred to as the winglet correlation length, and it represents the characteristic distance over which winglet motions were correlated. The effects of this length scale of the grid on the length scales of the flow are discussed in Sec.~\ref{sec:length_scales}.  

Time can be nondimensionalized by the grid correlation timescale, such that nondimensional time $t^* = t U/l_{\mathrm{grid}}$.

The difference in shape of these three autocorrelations can be quantified by defining a shape parameter
\begin{equation}
\zeta_{\mathrm{grid}} = -2\frac{\partial R_{\mathrm{grid}}}{\partial t^*}\Big|_{t^*=\epsilon},
\end{equation}
where $\epsilon$ indicates that the slope of the derivative is evaluated at a small nondimensional time. Physically, this shape parameter indicates how sharp the correlation function is at small correlation times and how broad the tail of the correlation is at large correlation times. When comparing two correlations with the same $l_{\mathrm{grid}}$, the one with a larger shape parameter has a more diffuse correlation, and the other has a more compact correlation. A boxcar kernel has a linear autocorrelation, so $\zeta_{\mathrm{grid}} = 1$ for a boxcar autocorrelation. For long tails, $\zeta_{\mathrm{grid}}>1$. This shape parameter may not be a general scaling parameter for all possible kernels, but it quantifies important differences between correlations produced by our various long-tail and boxcar kernels.

The effects of the shape parameter on the shape of the velocity correlation functions of the turbulence 
produced downstream of the grid are discussed in Sec.~\ref{sec:velocity_correlation_functions}.  

We developed two types of codes: codes that fixed the projected area (blockage) of the grid over time, and codes that did not constrain it. We observed no differences in the statistics that we report between the flows produced by these codes, and thus used the simpler code, in which the blockage was not constrained. The average blockage was between 0.1 and 0.2 across all experiments (0 and 1 correspond to unblocked and fully blocked, respectively). For any given experiment, the fluctuation in blockage was about 0.05.

\subsection{Instrumentation}
\label{sec:instrumentation}

In order to evaluate the degree to which correlated grid motions can influence turbulence properties, 
we collected time-series data of the flow velocity downstream of the active grid.  
Hot-wire probes were used to measure the flow velocity.  
A probe was mounted on two ISEL electronic traverses that traveled along the two spanwise dimensions of the tunnel.  
A custom traverse with a hand crank was used to move across the streamwise dimension of the tunnel.  
The data presented herein are taken with a single probe located at the center of the cross section of the tunnel, 
in order to avoid interference from the tunnel's boundary layers.  

Dantec single-wire and x-wire constant-temperature hot-wire anemometers were used in these experiments, 
controlled by a Dantec StreamLine system and digitally sampled at 20\,kHz.  
The hot wires were 5\,$\mu$m in diameter and 1\,mm long.  
Probe calibrations were performed adjacent to the experiment using a calibration unit 
that produced a laminar jet, according to methods described in \cite{browne,tropea}.  
To account for temperature fluctuations in the experimental hall, 
the temperature of the air exiting the tunnel was recorded at the beginning and end of every data set.  
The probe voltages were corrected using the relations described in \cite{bruun}, so that temperature variations of up to $\rm{3^o C}$ were taken into account.  

\subsection{Wind-tunnel setup and inflow conditions}
\label{sec:flow_properties}

While the active grid was designed for use in the VDTT described in Sec.~\ref{sec:mech_design}, the results presented in this paper were acquired with the grid mounted in an open-return wind tunnel built in 1935 
at the Kaiser-Wilhelm-Institut f{\"u}r Str{\"o}mungsforschung in G{\"o}ttingen run by L. Prandtl and H. Reichardt \cite[][]{Reichardt1938, Bodenschatz2011}.
The original test section of this tunnel was replaced with the present test section, which has the same cross-sectional profile as the VDTT test section.  
The dimensions are listed in the caption of Fig.\ \ref{fig:grid_lighted}.  
The 9.4-m test section was built of aluminum sheet metal mounted to a fire-proofed wooden structure 
as seen in Fig.\ \ref{fig:tunnel_schematic}.  
The wind-tunnel fan motor ran at a constant speed 
and produced a constant mean-flow velocity of approximately 1.45\,m/s. This velocity was employed mainly to ensure that the ratio of grid- to flow-response times $T^*$ remained below one.

The flow in the wind tunnel with the winglets of the active grid fully open 
was the baseline flow in the tunnel.  
Movements of the grid only added to this baseline turbulence.  
We measured the velocity at the center of the tunnel cross-section, 5.6\,m downstream of the grid. This location in the middle of the test section was chosen to mitigate transient effects from the grid and influences from the exit of the open wind tunnel; it also was the distance from the grid at which the turbulence produced by the grid was most isotropic (cf.\ Fig.\ \ref{fig:isotropy}). 
From these data we calculated the variance in the velocity signal to be $5.9 \times 10^{-4}$\,$\mathrm{m^2/s^2}$, 
so that the turbulence intensity was $1.7\%$.  
The mean flow speed was homogeneous to within about 2.5\% 
and the turbulence intensity to within about 5\% over the middle 75\% of the cross section of the tunnel.   
As in \cite{bodenschatz}, the production due to shear is compared to dissipation by examining the quantity 
\begin{equation}
\frac{\mathrm{production}}{\mathrm{dissipation}} = \frac{\langle uv \rangle \frac{\partial U}{\partial y} L}{u^{\prime 3}},
\end{equation}
where L is the integral length scale of the flow.
We find that this ratio equals $0.077 \pm 0.06$. These data suggest that production is small compared to dissipation yet larger than was observed in \cite{bodenschatz}.

\section{Results}
\label{Results}

In this section, we show how the grid control protocols described in Sec.~\ref{sec:grid_correlations} 
can be employed to control various turbulence parameters.  
The length scale, turbulence intensity, and Reynolds number can be precisely controlled.  
The shape of the velocity correlation function can be controlled 
by varying the shape of the grid correlation function.  
Such control is unprecedented in the literature.  
The grid affords some control over the isotropy and anisotropy of the flow, 
which is consistent with the findings in \cite{bewley}.  
Finally, we note that the changes in the turbulence structure did not have a significant 
impact on the decay rate of turbulent kinetic energy.  

\subsection{Turbulence intensity}
\label{sec:turbulence_intensity}

The turbulence intensity decays with distance from the grid (see Sec.~\ref{sec:energy_decay} for details). At a fixed distance from the grid, it is possible to vary the turbulence intensity by means of two mechanisms. The first is by increasing the fluctuations of the winglet angles. The second is by increasing the grid correlation length, thereby increasing the integral length scale of the turbulence (see Sec.~\ref{sec:length_scales}). Since turbulent kinetic energy decays with distance from grid normalized by the integral length scale, a larger integral length scale implies a higher turbulence intensity at a fixed location. 

\begin{figure}
	\includegraphics[width=\columnwidth]{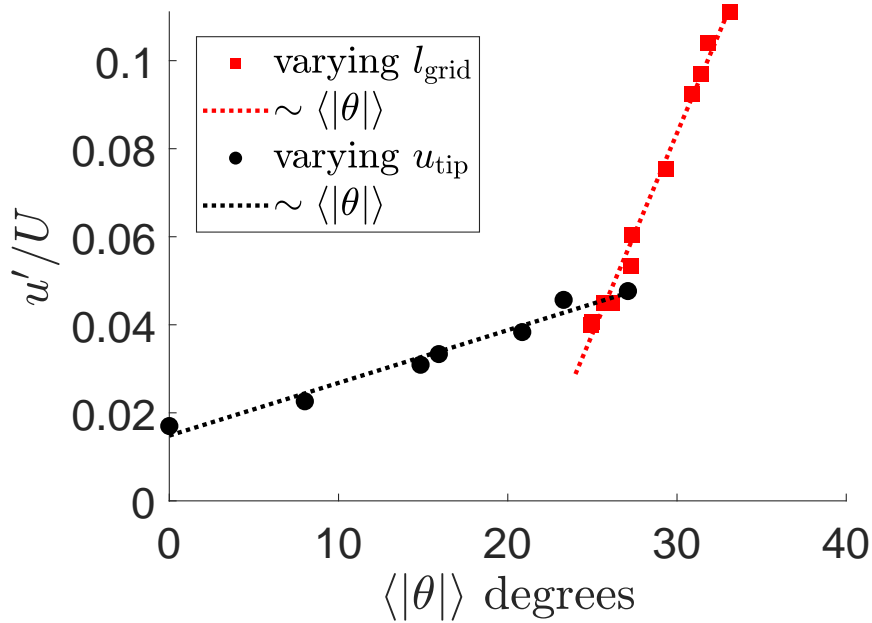}
    \caption{The turbulence intensity, $u^\prime / U$, is plotted against $\langle|\theta|\rangle$, the average winglet amplitude. $\langle|\theta|\rangle$ is varied through two mechanisms: first, by varying the tip speed of the winglets, $u_{\mathrm{tip}}$, from rest to 0.45$U$ (shown in black). Second, when the correlation length of the flow is increased, the effective decay distance of the turbulence decreases, so at a fixed location, the turbulence intensity increases (shown in red).}
    \label{fig:rmsvelpaddle}
\end{figure}

At a distance from the grid of 5.6\,m, the active grid could be used to vary the turbulence intensity in the tunnel 
between 1.7\% (stationary winglets) and 12\%, 
as shown in Fig.\ \ref{fig:rmsvelpaddle}.

To produce Fig.\ \ref{fig:rmsvelpaddle}, 
we collected 10 minutes of time-resolved velocity data for each point.
For a given protocol, when the RMS amplitude of the grid's winglet motions was increased, 
the magnitude of the RMS velocity fluctuations of the turbulence downstream of the grid increased monotonically. Since the average angle was zero, the RMS amplitude of the grid's winglet motions, and thus the magnitude of the tangential velocity of the tip of the winglets $u_{\mathrm{tip}}$, scaled with $\langle|\theta|\rangle$. This behavior is shown in the black data in Fig.\ \ref{fig:rmsvelpaddle}, and is consistent with the capabilities of other active grids \cite[e.g.][]{cekli}. For motions with relatively small amplitudes, the grid is essentially a classical grid with low solidity. For larger motions, the activity of the grid dominates. This effect is further shown in the red data in Fig.\ \ref{fig:rmsvelpaddle}, in which the RMS amplitude of the winglet motions was held constant and the length scale of the grid correlations was increased in both the streamwise and normal directions. The mechanism for this increase in turbulence intensity is that as the correlation length scale is increased, the length scale over which the turbulent kinetic energy decays is increased; thus, at a fixed location, the turbulence intensity increases as $l_{\mathrm{grid}}$ increases. In summary, Fig.\ \ref{fig:rmsvelpaddle} shows that the turbulence intensity can be varied directly by changing $u_{\mathrm{tip}}$ or indirectly using $l_{\mathrm{grid}}$.

\subsection{Length scales of the turbulence}
\label{sec:length_scales}

We begin by explaining how we computed the integral length scale
and the Taylor length scale.  
We then explain how we controlled each length scale using the active grid.

\subsubsection{Definition of the integral length scale} \label{sec:def_scales}
The integral length scale, $L$, is a measure of the largest scales of the turbulence.  
$L$ is defined as the integral under the velocity correlation function $R_{uu}(r)$, 
which was obtained by mapping temporal correlations in single-wire velocity measurements 
to spatial correlations using Taylor's Hypothesis.  
Since the velocity correlation function was noisy at large correlation distances, 
a numerical integration of the data was often unreliable at the tail of the correlation function.  
For this reason, an exponential best fit to the data was applied 
in the neighborhood of $R_{uu}(r) = 1/\mathrm{e}$.  
Beyond this point, the integral under the fit, instead of the integral under the noisy data, was computed.  
Linear fits were also considered, but the exponential fit was chosen since it proved more robust to noise.  

Another way to compute the length scale of large structures in the flow \cite[e.g.][]{tritton,bewley} 
is the distance $L_{1/\mathrm{e}}$ at which the velocity correlation $R_{uu}(L_{1/\mathrm{e}}) = 1/\mathrm{e}$.  
We computed both $L$ and $L_{1/\mathrm{e}}$ and found very good agreement, so we present $L_{1/\mathrm{e}}$ since there is no fitting or extrapolation required. We henceforth refer to $L_{1/\mathrm{e}}$ as $L$.

\subsubsection{Integral length scale variation}
\begin{figure}
    \includegraphics[width=\columnwidth]{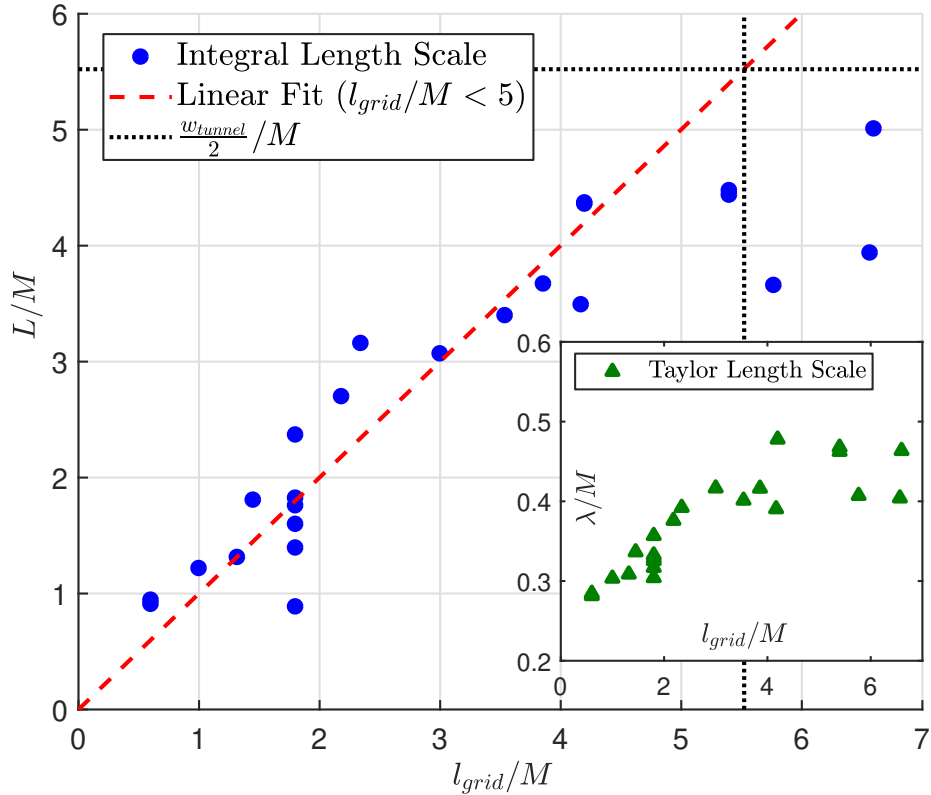}
    \caption{The integral length scale, $L/M$, is plotted in blue against the characteristic correlation length of the winglet forcing protocol, $l_{\mathrm{grid}}/M$. For small $l_{\mathrm{grid}}/M$, the correlations are governed by the size of each winglet in the grid. For large $l_{\mathrm{grid}}/M$, the correlations are limited by boundary-layer effects from the size of the tunnel (denoted on the plot as $\frac{w_{\mathrm{tunnel}}}{2}$, where $w_{\mathrm{tunnel}}$ is the tunnel's minimum diameter). The Taylor length scales for the same data, $\lambda/M$, are shown in green on an inset figure, showing that $\lambda$ follows a similar trend as $L$ ($u^{\prime}$ is not fixed so $\lambda^2 \not\sim L$).}
    \label{fig:Lintegral_vs_lgrid}
\end{figure}

To demonstrate that our active grid can control the integral length scale, 
many different grid forcing protocols were tested, corresponding to a wide space of $\zeta_{\mathrm{grid}}$ values (discussed in Sec.~\ref{sec:velocity_correlation_functions}). All of these protocols had matched streamwise and normal correlation lengths, so that an increase in $l_{\mathrm{grid}}$ implied proportional increases in the sizes of both streamwise and normal correlations. Independent of shape, the integral length scale varied monotonically with the length scale of the winglet correlations, $l_{\mathrm{grid}}$, 
as shown in Fig.\ \ref{fig:Lintegral_vs_lgrid}.  
Thus, the integral length scale could be directly controlled by changing the parameters that describe the grid motion. This relationship was limited by the minimum diameter of the test section.
This result is significant because it means that a single apparatus 
can generate turbulence at a range of different scales.  

\subsubsection{Taylor microscale variation}
The Taylor microscale, $\lambda$, 
is of interest since it is the length scale associated with the fluctuating rate of strain tensor and it is used to compute the turbulence Reynolds number, 
$Re_\lambda = {\lambda u^{\prime}}/{\nu}$.  
To calculate this scale, 
a concave-down parabola was fitted to $R_{uu}(r)$ for small $r$ ($0 \leq r \leq 0.01 L$), 
and the Taylor microscale was given by the r-intercept of this parabola.  

Like the integral length scale, the Taylor microscale varies monotonically with $l_{\mathrm{grid}}$ (shown in the inset of Fig.\ \ref{fig:Lintegral_vs_lgrid}), 
demonstrating that the kernel size could be used as an input parameter 
to control the magnitude of the Taylor microscale, 
and thereby control the turbulence Reynolds number in a systematic manner. 
This allowed us to keep the Reynolds number constant while changing other parameters about the grid motion.  
This capability is demonstrated in Fig.\ \ref{fig:velocity_correlation} 
where flows with vastly different correlation functions but similar Reynolds numbers are generated.

\subsection{Control of the velocity correlation functions}
\label{sec:velocity_correlation_functions}

Beyond controlling the size of generated turbulent structures, 
the active grid can control the shape of these structures.  
To assess this experimental capability, 
we conduct an investigation of the effect of shape of the grid motion correlations 
on the velocity correlation functions of the turbulence generated downstream of the grid.  
A series of long-tail and top-hat correlations were applied to the grid.  

\begin{figure}
    \includegraphics[width=\columnwidth]{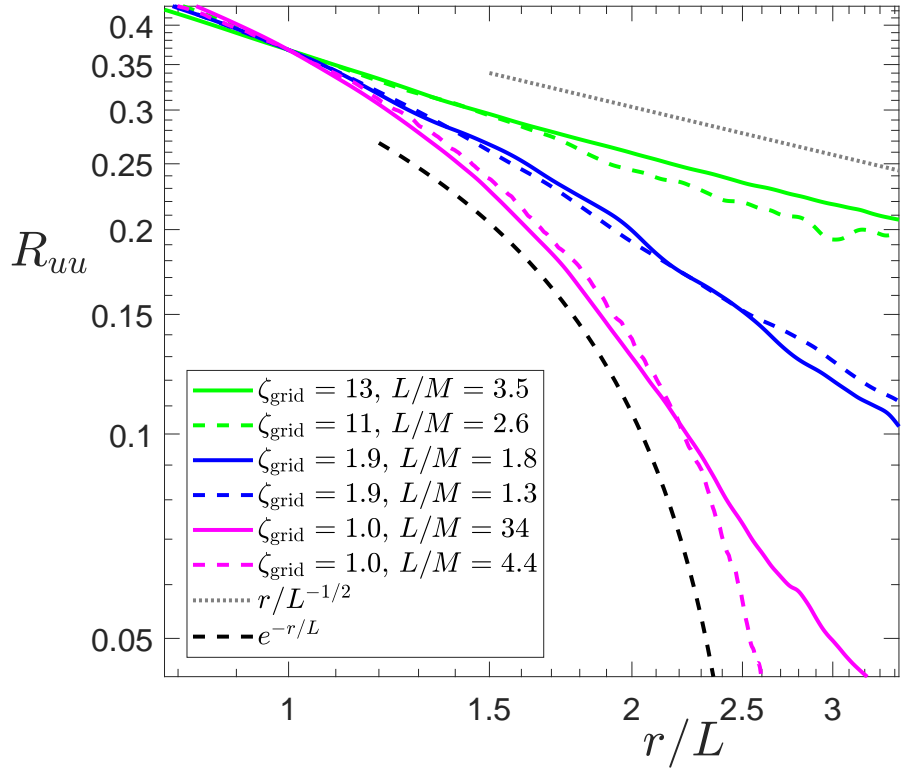}
    \caption{Velocity correlation functions for flows resulting from different winglet forcing protocols. 
    $Re_{\lambda}$ was kept approximately constant as the correlation shape and size were varied. Irrespective of $Re_{\lambda}$ and integral length scale $L/M$, the correlation functions normalized by integral length scale collapse into three groups, where each group corresponds to a different winglet correlation shape, ($\zeta_{\mathrm{grid}}$). The magenta forcings with $\zeta_{\mathrm{grid}}=1$ lead to exponential-like velocity correlations. The green forcings with the large $\zeta_{\mathrm{grid}}$ lead to power-law velocity correlations. The blue forcings of intermediate $\zeta_{\mathrm{grid}}$ lead to intermediate velocity correlations. In each group, there is one correlation function with smaller $L$ and one with larger $L$, showing that the changes to the velocity correlation functions are independent of length scale. For the green curves,  $Re_{\lambda} = 215 \pm 19$. For the blue curves, $Re_{\lambda} = 173 \pm 23$. For the magenta curves, $Re_{\lambda} = 225 \pm 10$.}
    \label{fig:velocity_correlation}
\end{figure}

Each velocity correlation function was computed from velocity measurements at the same location, 
sampled at 20\,kHz. 
In addition, leveraging the grid's control over $u^{\prime}$ and $\lambda$, the turbulence Reynolds number was held approximately constant (see caption of Fig.\ \ref{fig:velocity_correlation} for details).  
Two top-hat and four long-tail correlations are considered in Fig \ref{fig:velocity_correlation}.
The velocity correlation functions were normalized by each data set's integral length scale, 
in order to compare the shape and not the size of the velocity correlations. The $1/e$ definition of the integral scale (see Sec.~\ref{sec:def_scales}) is convenient for this purpose to ensure all correlations go through the point (1, $1/e$).

To parameterize the results in Fig.\ \ref{fig:velocity_correlation}, we define a shape parameter for the velocity correlations,
\begin{equation}
\zeta_{\mathrm{flow}} = \int \frac{r}{L} R_{uu}(r/L) d(r/L),
\end{equation}
which is the first moment of the velocity correlation function. Therefore, a larger value of $\zeta_{\mathrm{flow}}$ indicates a more diffuse correlation function (more power-law-like such as the green functions in Fig.\ \ref{fig:velocity_correlation}) and a smaller value of $\zeta_{\mathrm{flow}}$ indicates a more compact correlation function (more exponential such as the magenta curves in Fig.\ \ref{fig:velocity_correlation}). This interpretation of $\zeta_{\mathrm{flow}}$ for characterizing velocity correlations is consistent with the interpretation of $\zeta_{\mathrm{grid}}$ for interpreting winglet correlations (see Sec.~\ref{sec:grid_correlations}).

\begin{figure}
  \includegraphics[width=\columnwidth]{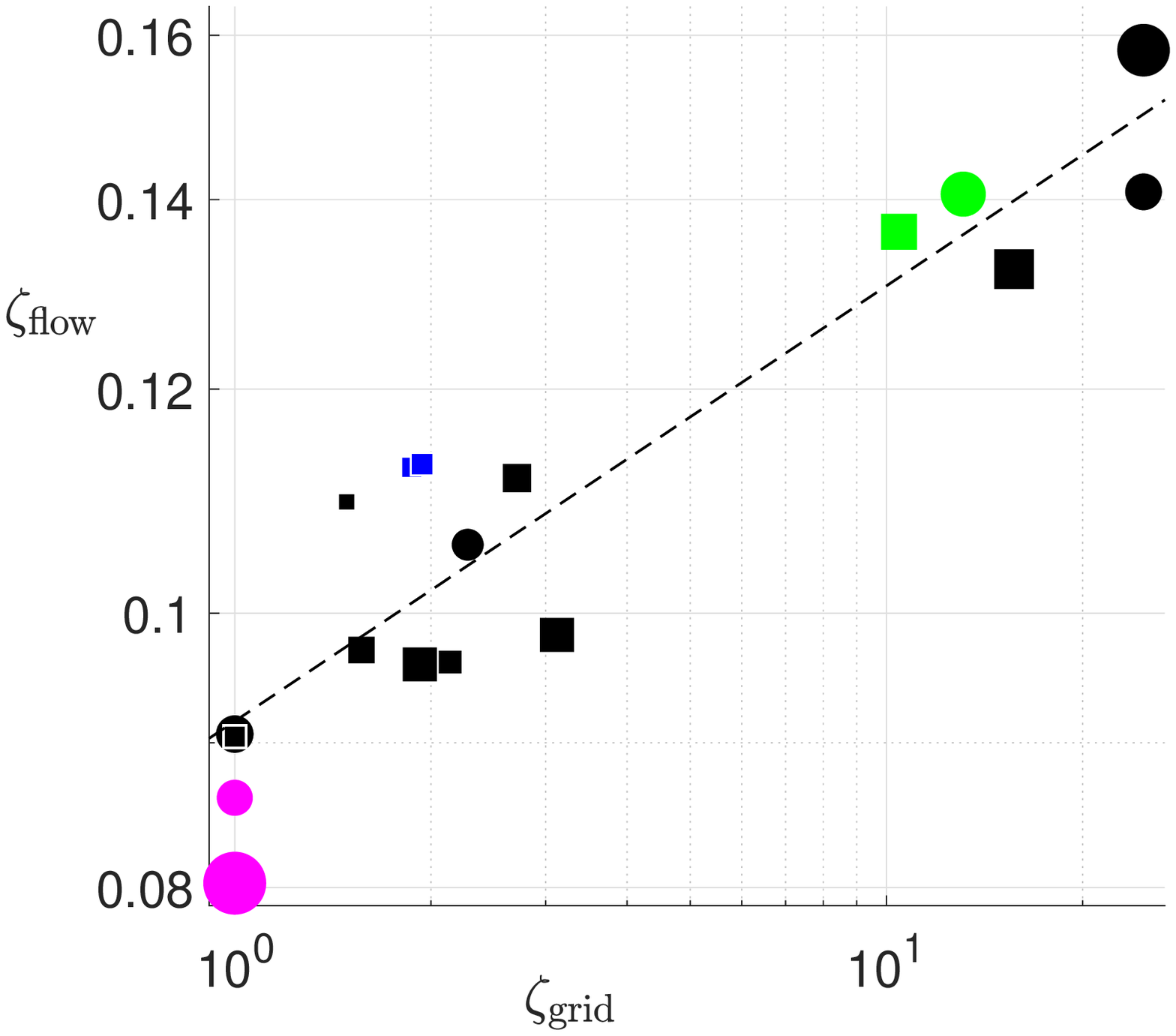}
  \caption{The shape of the velocity correlation, $\zeta_{\mathrm{flow}}$, is computed for various winglet forcing protocols and plotted against the shape of the winglet correlation, $\zeta_{\mathrm{grid}}$. The magenta, blue, and green data correspond to the velocity correlation data of the same color in Fig.\ \ref{fig:velocity_correlation}. The black data represent additional shapes and sizes of winglet correlations not shown in Fig.\ \ref{fig:velocity_correlation}. These data further support the conclusion that $\zeta_{\mathrm{grid}}$ is an experimental knob for controlling the velocity correlation function shape, $\zeta_{\mathrm{flow}}$. The size of the data represent the integral length scale. The smallest symbol corresponds to an integral scale of $L/M = 1.2$ and the largest symbol corresponds to an integral length scale of $L/M = 34$.} The fact that symbols of similar size are scattered uniformly demonstrates that the active grid is capable of controlling correlation-function shape independent of correlation-function size. A power law is drawn for reference as a dotted line. The squares represent data generated from 10-minute time series, and circles represent data from 60-minute time series. Longer samples are needed to increase the convergence in the square data.
  \label{fig:corr_vs_zeta}
\end{figure}

The winglet correlations tested were indeed three dimensional, but both $\zeta_{\mathrm{grid}}$ and $\zeta_{\mathrm{flow}}$ are parameterizations of a single dimension of the correlation functions. Thus, three-dimensional parameters may afford even more precise control of $\zeta_{\mathrm{flow}}$.

 Fig.\ \ref{fig:velocity_correlation} illustrates that the shape of the velocity correlation function downstream of the grid is determined by the grid control parameter $\zeta_{\mathrm{grid}}$, which the experimenter uses to control the shape of the winglet correlation. Irrespective of $L$ and of small changes in $Re_{\lambda}$, the velocity correlations fall into three groups: the magenta exponential-like correlations that have $\zeta_{\mathrm{grid}}=1$, the blue intermediate correlations that have intermediate $\zeta_{\mathrm{grid}}$, and the green power-law-like correlations that have large values of $\zeta_{\mathrm{grid}}$. The shape of the correlation (exponential-like or power-law-like) is quantified through the correlation shape parameter $\zeta_{\mathrm{flow}}$. Fig.\ \ref{fig:corr_vs_zeta} establishes that the three color groups that appeared in Fig.\ \ref{fig:velocity_correlation} indeed have similar shape. The additional data points in Fig.\ \ref{fig:corr_vs_zeta} (in black) further the claim that the grid control parameter $\zeta_{\mathrm{grid}}$ affords control of the velocity correlation shape.

As a brief summary, Fig.\ \ref{fig:Lintegral_vs_lgrid} illustrates that the experimenters can change the size of velocity correlations by varying the size of correlations on the grid, $l_{\mathrm{grid}}$. Fig.\ \ref{fig:velocity_correlation} and Fig.\ \ref{fig:corr_vs_zeta} establish that the experimenters can control the shape of the velocity correlations independent of the size of the correlations.

The results of these correlation-function experiments demonstrate that the active grid is able to change the structure of turbulence, particularly with regard to structure persistence and correlations at different scales, through the use of different types of grid-motion correlation functions.  
The ability to easily manipulate characteristics of structures in turbulent flow 
allows our active grid to more fully explore the the parameter space of various shapes and sizes of turbulent structures.

\subsection{Discussion of correlation shape control}

Note that spatial and temporal correlations are inherent to the construction of the grid.  
The algorithms we introduce add a degree of control over the extent of the correlations that are already present.  

A classical grid generates turbulence by shedding vortices from its constituent bars.  
The axes of these vortices are initially approximately aligned with the grid bars themselves.  
The vortices are carried downstream and interact with each other in a such a way 
that nearly uniform turbulence results some tens of grid spacings downstream of the grid \cite[][]{comte-bellot}.  
During the interaction of the vortices, the orientation of the grid is forgotten, 
but the grid spacing is imprinted in the correlation length of the flow.  

It is instructive to think that the active grid mixes the fluid in two ways.  
First, the winglets shed vortices from their upstream tips.  
The axes of these vortices are aligned with the mean flow direction.  
The vortices interact in a similar way as for the passive grid, 
and probably also yield a correlation length of the order of the winglet size.  
Second, as the winglets move to positions perpendicular to the flow, 
they block parts of the cross section of the tunnel.  
The evolving blockage of the cross section causes the flow to divert in different ways around the grid -- 
it is an obstacle of varying shape and size.  
If the motion of the winglets are coordinated in some way, 
correlated motions in the flow can be generated on scales larger than the individual winglets -- 
probably on the scale of the correlation length in the winglet positions.  

\subsection{Isotropy}
\label{sec:isotropy}

We hypothesized that changing the relative sizes of the spatial and temporal kernels used on the grid 
would correspond to similar changes in the isotropy of the generated flows.  
The isotropies of flows produced by kernels with different ratios of correlation lengths 
were analyzed using cross-wire probes at a series of locations along the length of the tunnel.  
Two grid correlations were used: 
one with matched streamwise and normal grid-correlation lengths, 
and one with a ratio of streamwise to normal correlation lengths of 1 to 12.5. This ratio is referred to as the anisotropy of the grid forcing, $I_{\mathrm{grid}}$.
Ten minutes of velocity data were collected at each tunnel location.  
At these locations, the kernel with equal streamwise and normal grid correlation lengths yielded $u^{\prime}/v^{\prime} \approx 1$, 
while the stretched kernel produced flows with $u^{\prime}/v^{\prime} > 1$.  

\begin{figure}
  \includegraphics[width=\columnwidth]{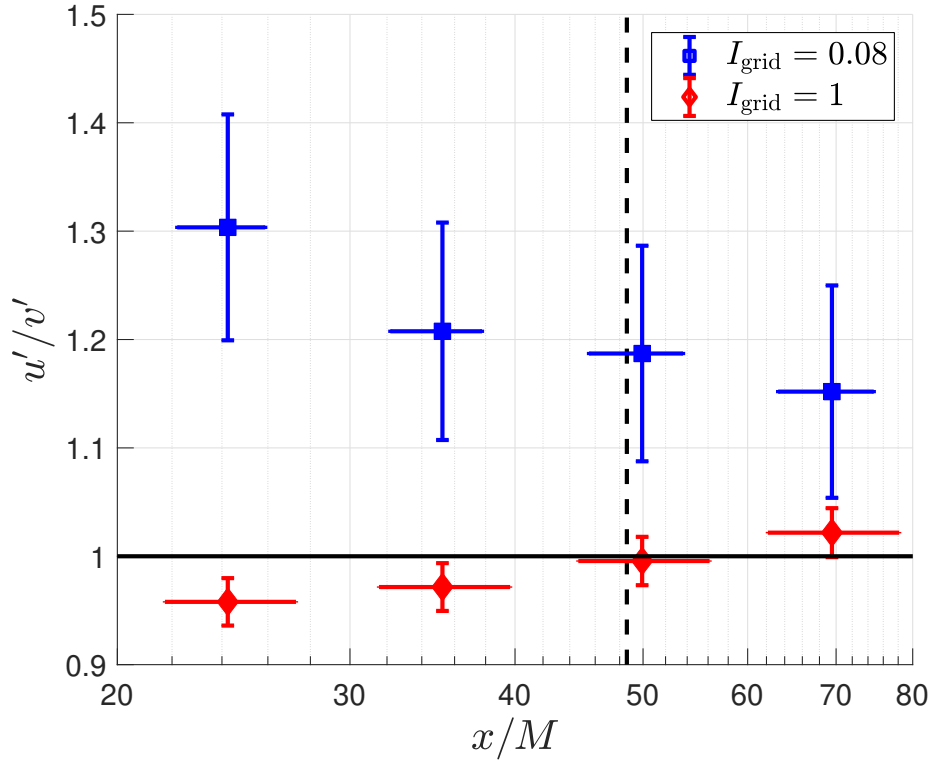}
  \caption{The isotropy of the flow along the tunnel was measured for two types of correlations. When the grid correlations were specified to be isotropic, the resulting flows were nearly isotropic as well. Conversely, highly anisotropic grid correlation protocols generated more anisotropic turbulent structures. The dotted line indicates the streamwise location at which the data for Fig.\ \ref{fig:rmsvelpaddle}-\ref{fig:corr_vs_zeta} were collected.}
  \label{fig:isotropy}
\end{figure}

The data are shown in Fig.\ \ref{fig:isotropy}.  
Because of the relatively small amount of data taken at each tunnel location, 
the errors in these measurements are significant, 
but the differences in $u^{\prime}/v^{\prime}$ observed still exceed the limits of these errors. Vertical error bars represent the standard error associated with sampling a known number of correlation lengths of data with known variance. Horizontal error bars are the result of averaging across a number of measurement positions to decrease the vertical error bars. 
The convergence of both of these data toward $u^{\prime}/v^{\prime} = 1$ 
at large distances reflects the tendency of decaying turbulence to move toward isotropy.  
Though not enough data were taken to extract clearly defined trends 
between kernel correlation lengths and flow isotropy, 
the initial experiments presented here demonstrate that the active grid 
allows the isotropy of the grid's turbulent structures to be altered in a systematic manner by manipulating the correlation lengths in the streamwise and transverse dimensions. This is consistent with the findings in \cite{bewley} and \cite{carter}.

\subsection{Decay of turbulent kinetic energy}
\label{sec:energy_decay}

An analysis of the decay behaviors of different kinds of turbulent structures 
using the grid's correlation kernel capabilities was a logical extension 
of the findings established thus far.  
Changes in the velocity correlation function imply changes in structure persistence in time, 
which intuitively should correspond to changes in turbulence decay rates.  
Preliminary tests of three representative kernels were carried out 
in order to investigate the effects of kernel shape on the normalized energy of turbulent fluctuations, 
${u^{\prime}}^2 / U^2$.  

\begin{figure}
	\includegraphics[width=1.0\columnwidth]{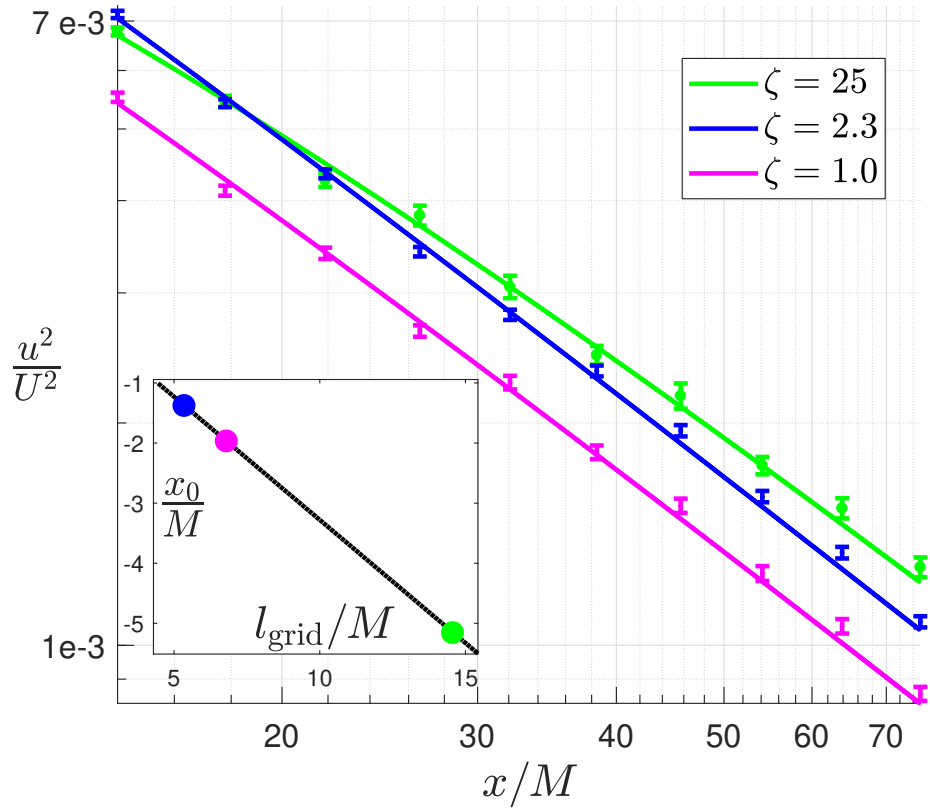}
    \caption{Three different grid correlation protocols, distinguished by different $\zeta_{\mathrm{grid}}$, were selected for energy decay tests. The energy of the velocity fluctuations was recorded at ten points along the tunnel. The energy decay profiles of the three grid correlation protocols had a similar shape despite the large differences in velocity correlation functions shown in Fig.\ \ref{fig:velocity_correlation}. The inset figure shows that the virtual origins ($x_0$) returned by the two-parameter power-law fits are linear with $l_{\mathrm{grid}}$. This suggests that the active grid may have a direct influence on the virtual origin of the power-law decay profiles. More data are needed to make this observation more specific.}
    \label{fig:energy_decay}
\end{figure}

Energy decay data is reported at 10 logarithmically spaced locations along the test section between 14 and 74 winglet-lengths downstream of the grid. Three winglet correlation protocols (with $\zeta$ values similar to those shown Fig.\ \ref{fig:paddle_corr}) are considered.
At each location, 80 minutes of data were taken for the long-tail correlation ($\zeta_{\mathrm{grid}} = 2.3$), 
90 minutes for the top-hat ($\zeta_{\mathrm{grid}} = 1$), 
and 160 minutes for the anisotropic long-tail ($\zeta_{\mathrm{grid}} = 25$).  
The varying time intervals of data collection were necessary to achieve the same standard error for the datasets with longer streamwise correlation lengths.  
When a three-parameter power-law fit is considered, the $95\%$ confidence intervals of the decay exponents overlapped, so that any difference in decay rate is not statistically significant. Therefore, a two-parameter power-law fit is considered.
In \cite{comte-bellot}, the turbulent kinetic energy is observed to decay according to the exponent 1.25.  
Similarly, \cite{Sinhuber2017} reports a decay exponent of 1.2.  
As the present apparatus has an octagonal test section identical to that of the latter reference, 
a decay exponent of 1.2 is considered.  
A two-parameter power-law fit is applied to the data to compute a virtual origin and scaling factor. The resulting fits are shown in Fig.\ \ref{fig:energy_decay}. In the corresponding inset, it is shown that the virtual origins of the fits vary linearly with the winglet parameter $l_{\mathrm{grid}}$. This suggests that the experimenters can precisely control the decay virtual origin by varying the correlation length of the winglet forcing protocol. Furthermore, no claim is made about the physical meaning of this virtual origin. Mathematically, the virtual origin of a two-parameter fit is the distance from the grid at which the kinetic energy best exhibits the assumed decay exponent (-1.2). Experimentally, the control knob $l_{\mathrm{grid}}$ may be varying either the underlying power law or the virtual origin. Either way, this active grid does appear to have some ability to affect the decay of kinetic energy behind the grid. More data are needed to support this conclusion.

\section{Conclusion}
\label{conclusion}

Because of its high number of independently controllable degrees of freedom, 
the new active grid that we present in this paper affords unprecedented control over the characteristics of turbulence.  
The length scales of the large-scale structures in the flow 
are directly affected by the correlation lengths of the correlations in the motions of the grid.  
The grid's control over the magnitude of turbulent velocity fluctuations 
means that the turbulence Reynolds number is manipulated simply by changing the motion of the grid. Independent control of turbulence intensity and length scale affords independent control of Reynolds number and length scale.
Finally, the grid exhibits the capacity to modify not only the correlation length scales of the turbulent flow 
but also the shape of the velocity correlation functions.  Early results indicate the ability of the apparatus to control the anisotropy of the turbulence and the energy decay behavior but these results are more preliminary than those aforementioned.

Our initial experimental results demonstrate the potential of this novel type of active grid 
to produce a large parameter space of flows from a single apparatus.  
Access to such a large parameter space enables an experimenter to match a wide range of turbulent conditions, 
such as those characteristic of boundary layers, jets, or atmospheric phenomena.


\begin{acknowledgements}
Since its inception more than ten years ago, 
many assistants and colleagues have made this experiment possible.  
They include the staff of the Max Planck Institute for Dynamics and Self-Organization -- 
Joachim Hesse, Andreas Kopp, Artur Kubitzek, Ortwin Kurre, Andreas Renner, Udo Schminke 
and their colleagues in the machine and electronics shops.  
We thank Helmut Eckelmann and Holger Nobach for helping to 
put the Prandtl tunnel back into running condition.  
The project was largely advanced by undergraduates 
from the University of G\"ottingen and Princeton University -- 
Florian K\"ohler, who 
wrote parts of the final active-grid control code 
and helped to build the new test section for the Prandtl tunnel, 
and Jessie Liu and Horace Zhang, who
developed initial versions of our methods and collected preliminary data.  
We thank the Princeton International Internship Program for funding the internships of 
Jessie Liu, Horace Zhang, and authors Kevin Griffin and Nathan Wei.  
Visiting graduate-student assistants included Ergun Cekli, 
who wrote the first active grid control code 
and helped to build the active grid, 
and Florent Lachauss{\'e}e, who developed new methods to switch the grid between different states.  
Finally, we acknowledge Zellman Warhaft for stimulating initial discussions about active grid design
and for providing access to his wind tunnel in which we measured the torque on winglets,  
Greg Voth who worked with Florent to acquire initial data, 
and Willem van de Water for his assistance throughout the project.  
\end{acknowledgements}

\bibliographystyle{spbasic}      
\bibliography{active_grid}   

\begin{thebibliography}{51}
\providecommand{\natexlab}[1]{#1}
\providecommand{\url}[1]{{#1}}
\providecommand{\urlprefix}{URL }
\expandafter\ifx\csname urlstyle\endcsname\relax
  \providecommand{\doi}[1]{DOI~\discretionary{}{}{}#1}\else
  \providecommand{\doi}{DOI~\discretionary{}{}{}\begingroup
  \urlstyle{rm}\Url}\fi
\providecommand{\eprint}[2][]{\url{#2}}

\bibitem[{Ayyalasomayajula et~al(2006)Ayyalasomayajula, Gylfason, Collins,
  Bodenschatz, and Warhaft}]{Ayyalasomayajula2006}
Ayyalasomayajula S, Gylfason A, Collins LR, Bodenschatz E, Warhaft Z (2006)
  {Lagrangian measurements of inertial particle accelerations in grid generated
  wind tunnel turbulence}. Phys Rev Lett 97(14):1--4

\bibitem[{Bewley et~al(2012)Bewley, Chang, and Bodenschatz}]{bewley}
Bewley GP, Chang K, Bodenschatz E (2012) {On integral length scales in
  anisotropic turbulence}. Phys Fluids 24(061702)

\bibitem[{Blum et~al(2011)Blum, Bewley, Bodenschatz, Gibert, {\'{A}}rmann,
  Mydlarski, Voth, Xu, and Yeung}]{Blum2011}
Blum DB, Bewley GP, Bodenschatz E, Gibert M, {\'{A}}rmann G, Mydlarski L, Voth
  GA, Xu H, Yeung PK (2011) {Signatures of non-universal large scales in
  conditional structure functions from various turbulent flows}. New J Phys 13

\bibitem[{Bodenschatz and Eckert(2011)}]{Bodenschatz2011}
Bodenschatz E, Eckert M (2011) {Prandtl and the G{\"o}ttingen school}. In:
  Davidson PA, Kaneda Y, Moffatt K, Sreenivasan KR (eds) A Voyag. Through
  Turbul., Cambridge University Press, pp 40--100

\bibitem[{Bodenschatz et~al(2014)Bodenschatz, Bewley, Nobach, Sinhuber, and
  Xu}]{bodenschatz}
Bodenschatz E, Bewley GP, Nobach H, Sinhuber M, Xu H (2014) {Variable density
  turbulence tunnel facility}. Rev Sci Instrum 85(9)

\bibitem[{Browne et~al(1989)Browne, Antonia, and Chua}]{browne}
Browne LWB, Antonia RA, Chua LP (1989) {Calibration of X-probes for turbulent
  flow measurements}. Exp Fluids 7

\bibitem[{Bruun(1995)}]{bruun}
Bruun HH (1995) {Hot-wire anemometry-principles and signal analysis}. Oxford
  Sci Publ

\bibitem[{Cal et~al(2010)Cal, Lebr{\'{o}}n, Castillo, Kang, and
  Meneveau}]{Cal2010}
Cal RB, Lebr{\'{o}}n J, Castillo L, Kang HS, Meneveau C (2010) {Experimental
  study of the horizontally averaged flow structure in a model wind-turbine
  array boundary layer}. J Renew Sustain Energy 2(1), \doi{10.1063/1.3289735}

\bibitem[{Carter and Coletti(2017)}]{carter}
Carter DW, Coletti F (2017) {Scale-to-scale anisotropy in homogeneous
  turbulence}. J Fluid Mech 827:250--284

\bibitem[{Cekli and van~de Water(2010)}]{cekli2}
Cekli HE, van~de Water W (2010) {Tailoring turbulence with an active grid}. Exp
  Fluids 49:409--416

\bibitem[{Cekli et~al(2010)Cekli, Tipton, and van~de Water}]{cekli3}
Cekli HE, Tipton C, van~de Water W (2010) {Resonant Enhancement of Turbulent
  Energy Dissipation}. Phys Rev Lett 105(4)

\bibitem[{Cekli et~al(2015)Cekli, Joosten, and van~de Water}]{cekli}
Cekli HE, Joosten R, van~de Water W (2015) {Stirring turbulence with
  turbulence}. Phys Fluids 27

\bibitem[{Chang et~al(2012)Chang, Bewley, and Bodenschatz}]{chang}
Chang K, Bewley GP, Bodenschatz E (2012) {Experimental study of the influence
  of anisotropy on the inertial scales of turbulence}. J Fluid Mech 692(464)

\bibitem[{Comte-Bellot and Corrsin(1971)}]{comte-bellot}
Comte-Bellot G, Corrsin S (1971) {Simple Eulerian time correlation in full and
  narrow band velocity signals in grid-generated, isotropic turbulence}. J
  Fluid Mech 48(2):273--337

\bibitem[{Corrsin(1942)}]{Corrsin1942}
Corrsin S (1942) {Decay of turbulence behind three similar grids}. PhD thesis,
  California Institute of Technology

\bibitem[{Garg and Warhaft(1998)}]{garg}
Garg S, Warhaft Z (1998) {On the small structure of simple shear flow}. Phys
  Fluids 10(3)

\bibitem[{Gerashchenko and Warhaft(2013)}]{Gerashchenko2013}
Gerashchenko S, Warhaft Z (2013) {Conditional entrainment statistics of
  inertial particles across shearless turbulent interfaces}. Exp Fluids 54(12),
  \doi{10.1007/s00348-013-1631-2}

\bibitem[{Gylfason and Warhaft(2004)}]{Gylfason2004}
Gylfason A, Warhaft Z (2004) {On higher order passive scalar structure
  functions in grid turbulence}. Phys Fluids 16(11):4012--4019,
  \doi{10.1063/1.1790472}

\bibitem[{Gad-el Hak and Corrsin(1974)}]{Gad-el-Hak1974}
Gad-el Hak M, Corrsin S (1974) {Measurements of the nearly isotropic turbulence
  behind a uniform jet grid}. J Fluid Mech 62:115--143

\bibitem[{Hearst and Ganapathisubramani(2017)}]{Hearst2017}
Hearst RJ, Ganapathisubramani B (2017) ailoring incoming shear and turbulence
  profiles for lab-scale wind turbines. Wind Energy 20(August):2021--2035,
  \doi{10.1002/we.2138}

\bibitem[{Hearst and Lavoie(2015)}]{hearst}
Hearst RJ, Lavoie P (2015) {The effect of active grid initial conditions on
  high Reynolds number turbulence}. Exp Fluids 56(185)

\bibitem[{Kang and Meneveau(2008)}]{kang2}
Kang HS, Meneveau C (2008) {Experimental study of an active grid-generated
  shearless mixing layer and comparisons with large-eddy simulation}. Phys
  Fluids 20(125102)

\bibitem[{Kang et~al(2003)Kang, Chester, and Meneveau}]{kang}
Kang HS, Chester S, Meneveau C (2003) {Decaying turbulence in an
  active-grid-generated flow and comparisons with large-eddy simulation}. J
  Fluid Mech 480:129--160

\bibitem[{Kastrinakis and Eckelmann(1983)}]{Eckelmann1983}
Kastrinakis EG, Eckelmann H (1983) {Measurement Of Streamwise Vorticity
  Fluctuations In A Turbulent Channel Flow}. J Fluid Mech 137:165--186

\bibitem[{Knebel and Peinke(2009)}]{knebel}
Knebel P, Peinke J (2009) {Active grid generated turbulence}. Adv Turbul XII
  Springer Proc Phys 132:903

\bibitem[{Knebel et~al(2011)Knebel, Kittel, and Peinke}]{knebel2}
Knebel P, Kittel A, Peinke J (2011) {Atmospheric wind field conditions
  generated by active grids}. Exp Fluids 51:471--481

\bibitem[{Kolmogorov(1941)}]{Kolmogorov1941b}
Kolmogorov AN (1941) {The local structure of turbulence in incompressible
  viscous fluid for very large Reynolds numbers}. Dokl Akad Nauk SSSR
  30(4):9--13, \doi{10.1098/rspa.1991.0075}

\bibitem[{Larssen and Devenport(2011)}]{Larssen2011}
Larssen JV, Devenport WJ (2011) {On the generation of large-scale homogeneous
  turbulence}. Exp Fluids 50(5):1207--1223, \doi{10.1007/s00348-010-0974-1}

\bibitem[{Makita(1991)}]{makita}
Makita H (1991) {Realization of a large-scale turbulence field in a small wind
  tunnel}. Fluid Dyn Res 8(53)

\bibitem[{Maldonado et~al(2015)Maldonado, Castillo, Thormann, and
  Meneveau}]{Maldonado2015}
Maldonado V, Castillo L, Thormann A, Meneveau C (2015) {The role of free stream
  turbulence with large integral scale on the aerodynamic performance of an
  experimental low Reynolds number S809 wind turbine blade}. J Wind Eng Ind
  Aerodyn 142:246--257

\bibitem[{Mathieu and Alcaraz(1965)}]{Mathieu1965}
Mathieu J, Alcaraz E (1965) {R{\'{e}}alisation d'une soufflerie {\`{a}} haut
  niveau de turbulence}. CR Acad Sci 261(2435)

\bibitem[{Mydlarski(2017)}]{mydlarski_review}
Mydlarski LB (2017) {A turbulent quarter century of active grids: From {Makita}
  (1991) to the present}. Fluid Dyn Res

\bibitem[{Mydlarski and Warhaft(1996)}]{mydlarski}
Mydlarski LB, Warhaft Z (1996) {On the onset of high-Reynolds-number
  grid-generated wind tunnel turbulence}. J Fluid Mech 320

\bibitem[{Obligado et~al(2015)Obligado, Cartellier, and
  Bourgoin}]{Obligado2015}
Obligado M, Cartellier A, Bourgoin M (2015) {Experimental detection of
  superclusters of water droplets in homogeneous isotropic turbulence}. Epl
  112(5)

\bibitem[{Poorte and Biesheuvel(2002)}]{poorte}
Poorte REG, Biesheuvel A (2002) {Experiments on the motion of gas bubbles in
  turbulence generated by an active grid}. J Fluid Mech 461

\bibitem[{Reichardt(1938)}]{Reichardt1938}
Reichardt VH (1938) {Messungen turbulenter Schwankungen}. Naturwissenschaften
  26(24/25):404--408

\bibitem[{Saw et~al(2012)Saw, Shaw, Salazar, and Collins}]{Saw2012}
Saw EW, Shaw RA, Salazar JP, Collins LR (2012) {Spatial clustering of
  polydisperse inertial particles in turbulence: II. Comparing simulation with
  experiment}. New J Phys 14(1991)

\bibitem[{Shen and Warhaft(2000)}]{shen1}
Shen X, Warhaft Z (2000) {The anisotropy of the small scale structure in high
  Reynolds number turbulent shear flow}. Phys Fluids 12(2976)

\bibitem[{Siebert et~al(2010)Siebert, Gerashchenko, Gylfason, Lehmann, Collins,
  Shaw, and Warhaft}]{Siebert2010}
Siebert H, Gerashchenko S, Gylfason A, Lehmann K, Collins LR, Shaw RA, Warhaft
  Z (2010) {Towards understanding the role of turbulence on droplets in clouds:
  In situ and laboratory measurements}. Atmos Res 97(4):426--437

\bibitem[{Sinhuber et~al(2017)Sinhuber, Bewley, and Bodenschatz}]{Sinhuber2017}
Sinhuber M, Bewley GP, Bodenschatz E (2017) {Dissipative Effects on
  Inertial-Range Statistics at High Reynolds Numbers}. Phys Rev Lett
  119(13):1--5

\bibitem[{Szasz{\'{a}}k et~al(2018)Szasz{\'{a}}k, Roloff, Bord{\'{a}}s, Bencs,
  Szab{\'{o}}, and Th{\'{e}}venin}]{Szaszak}
Szasz{\'{a}}k N, Roloff C, Bord{\'{a}}s R, Bencs P, Szab{\'{o}} S,
  Th{\'{e}}venin D (2018) {A novel type of semi-active jet turbulence grid}.
  Heliyon 4, \doi{10.1016/j.heliyon.2018.e01026}

\bibitem[{Thormann and Meneveau(2014)}]{thormann}
Thormann A, Meneveau C (2014) {Decay of homogeneous, nearly isotropic
  turbulence behind active fractal grids}. Phys Fluids 26

\bibitem[{Thormann and Meneveau(2015)}]{thormann2}
Thormann A, Meneveau C (2015) {Decaying turbulence in the presence of a
  shearless uniform kinetic energy gradient}. J Turbul 16(5):442--459

\bibitem[{Tritton(1988)}]{tritton}
Tritton DJ (1988) {Physical fluid dynamics}. Oxford University Press

\bibitem[{Tropea et~al(2007)Tropea, Yarin, and Foss}]{tropea}
Tropea C, Yarin AL, Foss JF (2007) {Springer handbook of experimental fluid
  mechanics}. Springer Science \& Business Media

\bibitem[{Varshney and Poddar(2011)}]{varshney}
Varshney K, Poddar K (2011) {Experiments on integral length scale control in
  atmospheric boundary layer wind tunnel}. Theor Appl Clim 106:127--137

\bibitem[{W{\"{a}}chter et~al(2012)W{\"{a}}chter, Hei{\ss}elmann,
  H{\"{o}}lling, Morales, Milan, M{\"{u}}cke, Peinke, Reinke, and
  Rinn}]{wachter}
W{\"{a}}chter M, Hei{\ss}elmann H, H{\"{o}}lling M, Morales A, Milan P,
  M{\"{u}}cke T, Peinke J, Reinke N, Rinn P (2012) {The turbulent nature of the
  atmospheric boundary layer and its impact on the wind energy conversion
  process}. J Turbul 13(26):1--21

\bibitem[{Warhaft and Shen(2002)}]{shen3}
Warhaft Z, Shen X (2002) {On the higher order mixed structure functions in
  laboratory shear flow}. Phys Fluids 14(2432)

\bibitem[{Weitemeyer et~al(2013)Weitemeyer, Reinke, Peinke, and
  H{\"{o}}lling}]{weitemeyer}
Weitemeyer S, Reinke N, Peinke J, H{\"{o}}lling M (2013) {Multi-scale
  generation of turbulence with fractal grids and an active grid}. Fluid Dyn
  Res 45(6)

\bibitem[{Yoon and Warhaft(1990)}]{yoon}
Yoon K, Warhaft Z (1990) {The evolution of grid-generated turbulence under
  conditions of stable thermal stratification}. J Fluid Mech 215:601--638

\bibitem[{Zhou and Venayagamoorthy(2019)}]{Zhou2019}
Zhou J, Venayagamoorthy SK (2019) {Near-field mean flow dynamics of a
  cylindrical canopy patch suspended in deep water}. J Fluid Mech 858:634--655,
  \doi{10.1017/jfm.2018.775}

\end{thebibliography}

\end{document}